\documentclass{aa}

\usepackage{graphicx}
\usepackage{txfonts}
\usepackage{lipsum}
\usepackage{subcaption}         
                               
\usepackage{lscape}            
                               
\usepackage{placeins}

\usepackage[colorlinks=true, citecolor=blue, linkcolor=blue, urlcolor=blue]{hyperref}

\usepackage{orcidlink}    

\newcommand{\src}{{\rm\,AT2021ehb}}

\newcommand{\msun}{M_{\odot}}
\newcommand{\xmm}{{XMM-Newton}}

\begin{document}

\title{Discovery of a soft X-ray lag in the tidal disruption event AT2021ehb}

\author{Wenjie Zhang\inst{1}\thanks{Corresponding author: zhangwj@bao.ac.cn}\orcidlink{0009-0003-9214-7316}}

\institute{National Astronomical Observatories, Chinese Academy of Sciences, Beijing 100101, China\\}

\date{Received December 9, 2025}

\abstract 
{In this Letter, we report the detection of soft X-ray time lags—i.e. variability in the softer photons lagging behind that in the harder photons—in seven \textit{XMM–Newton} observations of the tidal disruption event (TDE) candidate \src. We find correlated variability between the soft (0.3–0.7 keV) and hard (0.9–10 keV) bands on $\sim$$10^{4}$ s time-scales, and measure a soft lag of $\sim$500 s. This behaviour is broadly consistent with the disk–corona reverberation scenario established in active galactic nuclei (AGNs). Together with the previously reported strong hard X-ray emission and broad Fe K line, our results suggest the presence of a compact corona and prominent relativistic disk reflection in \src. The unusually high blackbody temperature (peaking at $\sim$200 eV) is difficult to reconcile with thermal emission from a standard accretion disk around a $\sim$$10^{7}M_{\odot}$ black hole, and may instead be analogous to the soft excess commonly observed in AGNs, whose physical origin remains debated. Finally, the measured lags offer a possible explanation for the rapid X-ray flux decline that occurred only three days after the peak, pointing to a scenario in which the corona cools following a sudden loss of the magnetic support required to sustain it.}


\maketitle
\nolinenumbers

\section{Introduction}

Rapid X-ray variability is often observed in active galactic nuclei (AGNs) and is generally understood to originate in the immediate vicinity of the central supermassive black hole (SMBH; \citet{Lynden-Bell1969,Mchardy2006,Kara2025}), thereby providing a powerful probe of the physical conditions and geometry of the innermost accretion flow. One particularly powerful application is X-ray reverberation mapping, which probes the disk–corona geometry through light-travel delays between the coronal continuum and the reprocessed emission. These X-ray lags trace regions within only a few gravitational radii, offering rare access to explore the structure of the immediate vicinity of the central black holes, and providing critical confirmation of the relativistic reflection model \citep{Fabian2009,Fabian2010,Zoghbi2010,Kara2013a,Kara2013b,Kara2013c,Uttley2014,Kara2016a}. Within this framework, the disk reflection naturally gives rise to the soft excess observed in many AGNs, through either blended low-ionization line emission or a bremsstrahlung-like continuum. This mechanism also provides a unified explanation for the short soft X-ray lags commonly associated with the soft excess. Nonetheless, this interpretation faces challenges, as fitting the soft excess often pushes reflection models toward extreme and potentially biased parameter regimes, suggesting that another mechanism
—a warm corona that comptonizes the UV photons from the inner accretion disk to soft X-ray~\citep{Czerny1987,Jin2009,Middleton2009,Done2012}—may also jointly contribute to producing the observed soft excess~\citep{Ballantyne2020,Xiang2022,Ballantyne2024}.
However, because AGNs are observed only in their already-developed states, the origin of the soft excess and the formation of the disk–corona structure remain difficult to constrain.

Transient accretion systems can form when a star passes sufficiently close to a SMBH and is torn apart once the tidal forces exceed its self-gravity, theoretically producing a sudden, significant X-ray brightening accompanied by an unusually soft, thermal X-ray spectrum~\citep{Rees1988}. Recent work suggests that, when sufficient magnetic flux is advected inward with the returning stellar debris, a new X-ray–emitting corona can form in TDEs~\citep{Begelman2014,Xu2025}. Observational evidence further supports this picture: several events have exhibited late-time hard X-ray emission ($>$2 keV) indicative of corona formation, enabled by the growing TDE sample and extensive X-ray follow-up campaigns across multiple observatories~\citep{Komossa2004,Kara2018,Jonker2020,Wevers2021,Yao2022,Guolo2024,Yao2024}. 
Furthermore, analyses of soft and hard X-ray variability in TDEs have recently begun, though to date soft X-ray lags have been reported in only two cases: Swift J1644+57~\citep{Kara2016b} and AT2018fyk~\citep{Zhangwd2022}. The former is a jetted TDE~\citep{Bloom2011,Levan2011,Zauderer2011}, and the detection of a soft lag indicates reverberation arising from deep within the relativistic potential, demonstrating that the X-rays do not originate from the relativistically moving regions of a jet~\citep{Kara2016b}. The latter underwent a transition from a disk-dominated to a corona-dominated state~\citep{Wevers2019,Wevers2021}, and the soft lags observed during its hard state are difficult to reconcile with the reverberation scenario typically invoked for AGNs. Instead, they are consistent with a picture in which the soft X-rays are produced by down-scattering of hard X-rays in an outflow~\citep{Zhangwd2022}.

AT2021ehb is a nearby TDE that exhibits unusually complex X-ray evolution over its first 430 days~\citep{Yao2022}. The host galaxy analysis indicates a central black hole mass of $M_{\rm BH}\sim10^{7}\msun$, whereas X-ray spectral modeling by \citet{Xiang2024} yields a much lower estimate of $\sim10^{5.5}\msun$. The event shows a delayed rise in X-ray luminosity, followed by a gradual transition from a soft, thermal spectrum to a hard, nonthermal one, and eventually a rapid return to a soft state accompanied by an order-of-magnitude flux drop. Joint \textit{NICER} and \textit{NuSTAR} observations reveal strong hard X-ray emission up to 30~keV and an extremely broad Fe K emission line, pointing to a compact corona and pronounced relativistic disk reflection. Despite these dramatic X-ray changes, the UV–optical emission remains comparatively steady and no radio emission is detected, suggesting a highly aspherical configuration without a relativistic jet~\citep{Dai2018,Alexander2020,Yao2022}. \citet{Yao2022} interpret the evolving X-ray properties as evidence for the gradual formation of a corona, anisotropic escape of hard X-rays, and a late-time disk instability that may temporarily modify the inner accretion flow geometry.
In this Letter, we analyze the \xmm\ data obtained after the sudden flux drop and perform a timing study to investigate the origin of the soft X-ray emission in \src. Using the measured lags, we further provide a plausible explanation for the abrupt decline in X-ray flux. In Section~2, we describe the observations and data reduction. Section~3 presents our X-ray timing and spectral analyses, and in Section~4 we discuss and summarize the implications of our results.

\begin{figure}[ht]
\centering
\includegraphics[width=\linewidth]{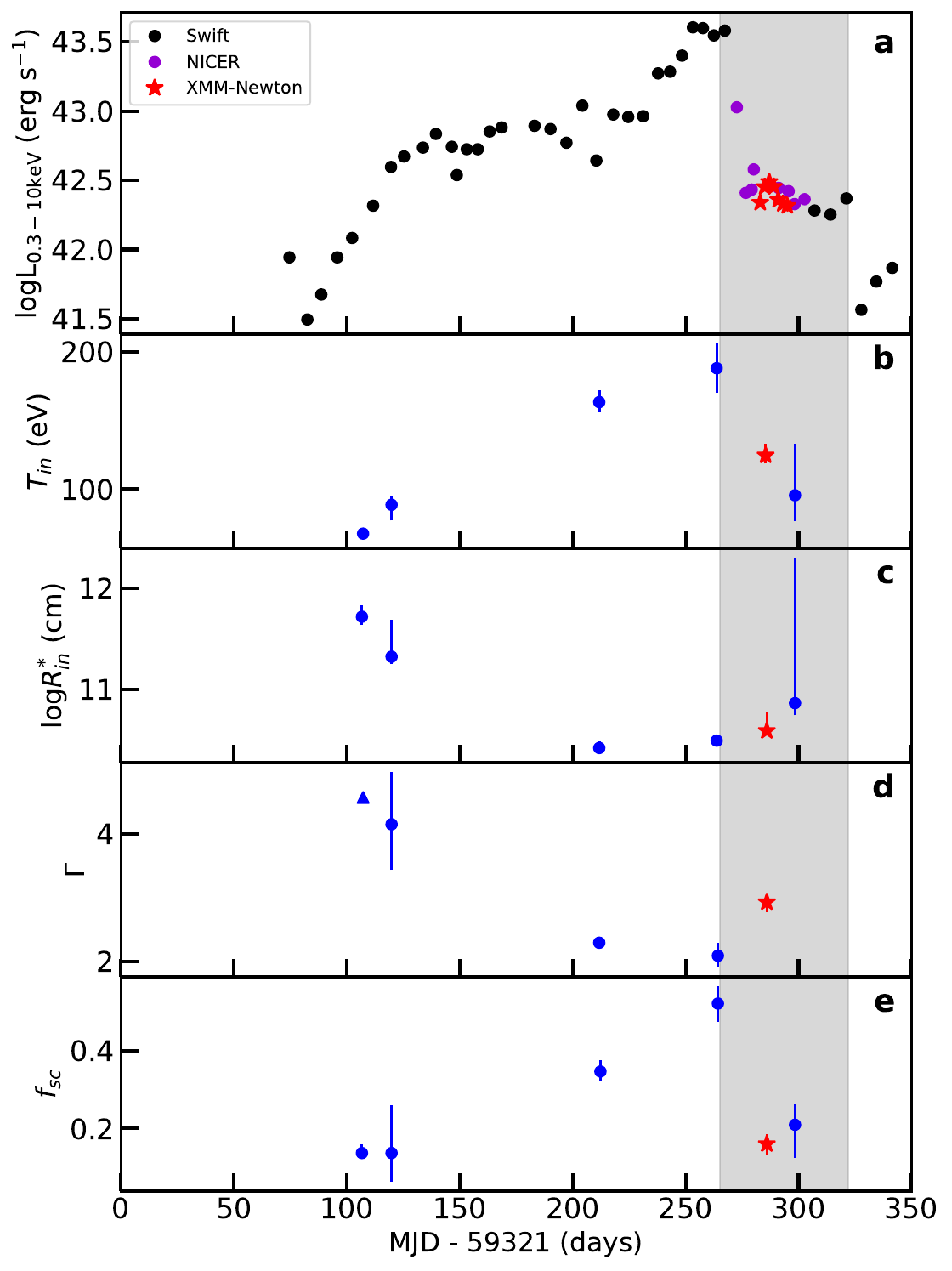}
\caption{Panel a: X-ray luminosity of \src\ in the 0.3–10 keV band.
Panels b–e show the evolution of the spectral parameters, where
$T_{\rm in}$ denotes the inner-disk temperature and $R_{\rm in}^{*}$ refers to the apparent inner-disk radius derived from the diskbb model, defined as the true inner-disk radius multiplied by $\sqrt{\cos i}$ (where $i$ is the disk inclination), while $\Gamma$ and $f_{\rm sc}$ represent the photon index of the comptonized power-law component and the fraction of seed photons scattered into it. In panels b–e, the red star denotes the mean values derived from the spectral fits to the seven \textit{XMM-Newton} observations, while the blue points are taken from \citet{Yao2022}.
}
\label{lc}
\end{figure}

\section{Observations and data reductions}

We obtained the long-term X-ray light curve and the evolution of the spectral parameters of \src\ directly from the measurements reported in \citet{Yao2022}. The light curve is primarily based on \textit{Swift}/XRT observations, with additional \textit{NICER} data incorporated to fill gaps in the XRT coverage (Fig.~\ref{lc}).

\src~was observed by \textit{XMM-Newton} at ten epochs. The first observation was largely affected by high particle background, and the last two observations yielded only marginal detections with insufficient photons for timing analysis. Therefore, we focus on the remaining seven observations, which were obtained between 2022-01-25 and 2022-02-06 with a cadence of approximately two days. We primarily used the data from the EPIC-PN (PN) camera, which has much higher sensitivity.
The data were processed with the Science Analysis Software (SAS) version~17.0.0 using the latest calibration files. 
Event files were extracted from circular regions with radii of 35$^{\prime\prime}$ centered on the source position determined from optical observations (R.A.=03:07:47.82, Dec.=+40:18:40.85; \citet{Yao2022}).
Background events were taken from four nearby source-free circular regions with radii of 35$^{\prime\prime}$. 
Photon pile-up was checked using the \texttt{epatplot} command, and was found to be negligible. The epochs of high-background events were examined using the light curves in the energy band 10-12 keV. We adopted a count-rate filtering criterion of 0.6~counts~s$^{-1}$ to remove time intervals affected by high particle background. 
We extracted the source spectra in the 0.3–10 keV range. For the timing analysis, we generated background-subtracted light curves in different energy bands, using identical start and end times, with the SAS task \textsc{epiclccorr} and a time bin size of 10 s.
Only good events with \texttt{PATTERN}~$\leq$~4 were used to generate the spectra and light curves.

\begin{figure}[ht]
\centering
\includegraphics[width=\linewidth]{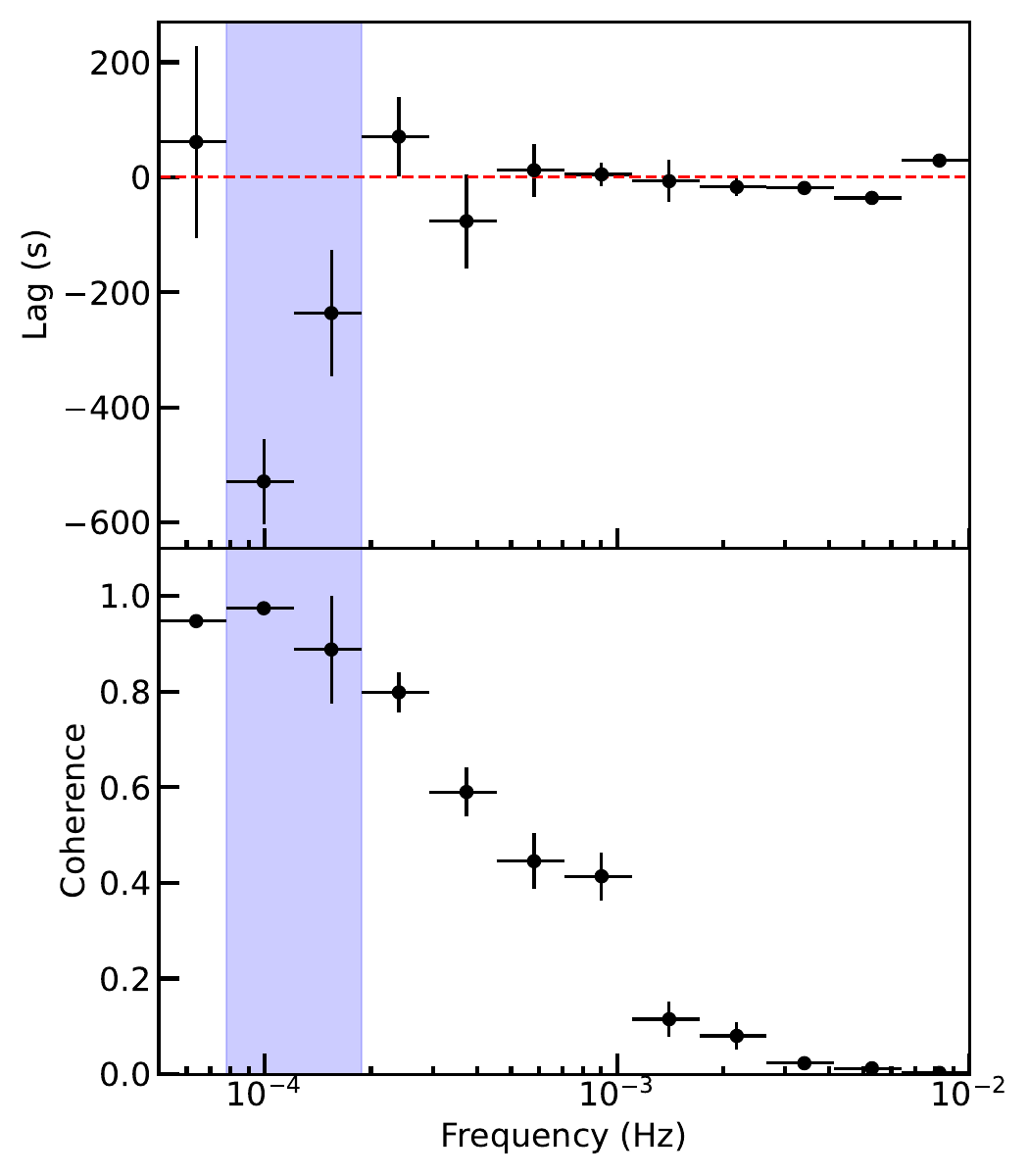}
\caption{Top panel: Soft (0.3–0.7 keV) versus hard (0.9–10 keV) lags as a function of temporal frequency. By convention, negative lags correspond to the soft band lagging behind the hard band. Variability on time-scales of 10$^{4}$ s shows a soft lag of approximately 500 s.
Bottom panel: Coherence as a function of frequency between the 0.3–0.7 and 0.9–10 keV bands. The blue-shaded region highlights the frequency range in which the soft band lags the hard band, during which the coherence remains close to 1. The blue-shaded frequency range is used to construct the energy–lag spectrum and the covariance spectrum. All statistical errors correspond to the 90\% confidence level.
}
\label{freq_lag}
\end{figure}

\begin{figure}[ht]
\centering
\includegraphics[width=\linewidth]{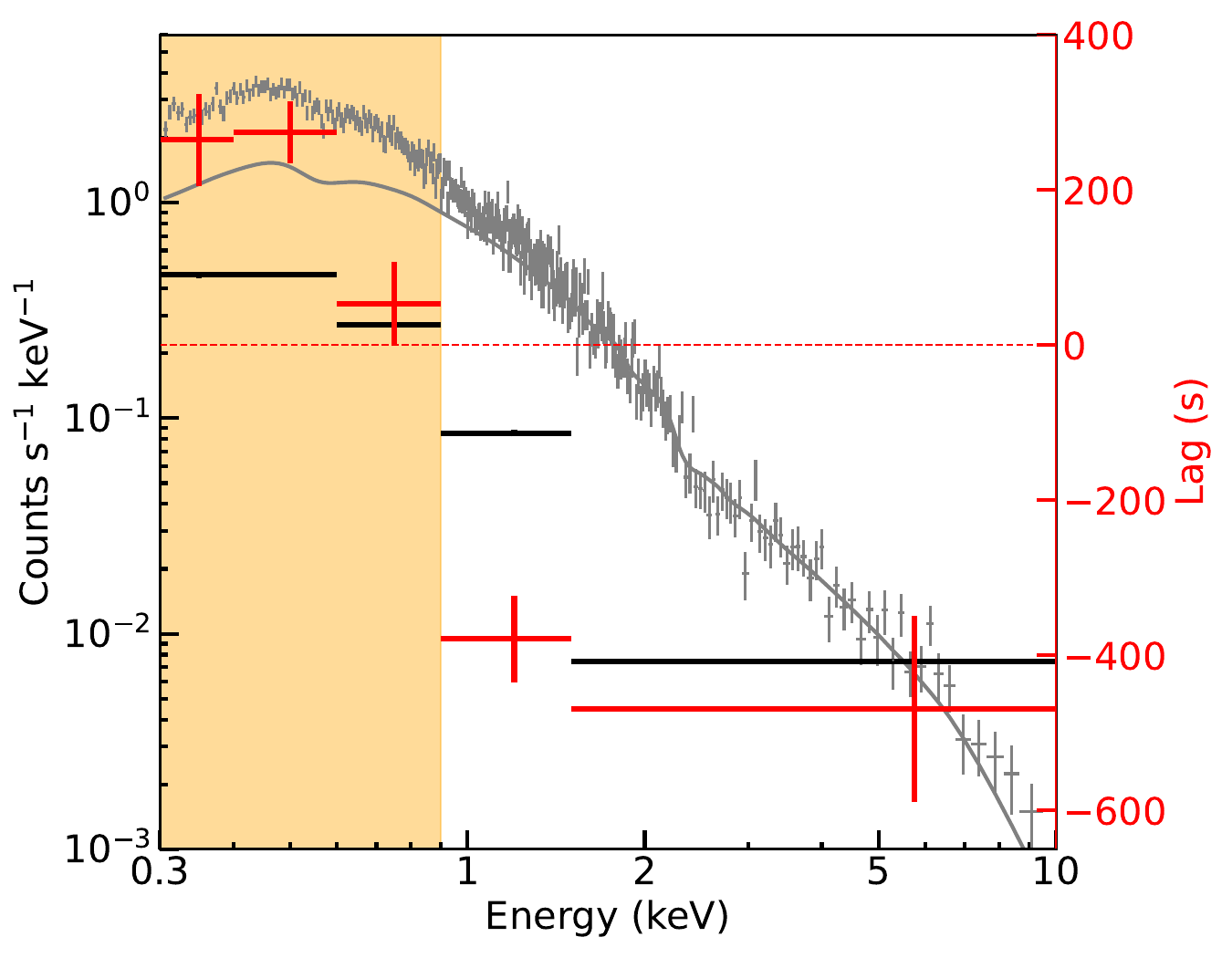}
\caption{Energy-dependent variability functions. The red points show the energy–lag spectrum, and the black points show the covariance spectrum. The grey points represent the time-averaged pn spectrum, while the best-fitting model with the \texttt{diskbb} component removed is shown as a gray curve. The orange-shaded region marks the energies at which the soft band lags the reference band, corresponding closely to the energy range where the \texttt{diskbb} component contributes significantly in the time-averaged spectrum. All statistical errors correspond to the 90\% confidence level.
}
\label{spectra}
\end{figure}

\section{Results}

A particularly striking feature visible in the Fig.~\ref{lc} panel is the rapid drop in X-ray luminosity (gray-shaded region), during which the flux declines by nearly an order of magnitude in only $\sim$3 days. All seven \textit{XMM-Newton} observations analyzed in this work were obtained after this rapid decline, each providing more than 18 ks of clean exposure and thereby enabling detailed timing analysis.
Benefiting from the fact that the seven \textit{XMM-Newton} observations were obtained over a short time span, we treat the first 18 ks of each observation as an individual segment. By processing seven segments in an identical manner and then averaging the resulting products, we obtain the coherence, frequency–lag spectra, energy–lag spectra, and covariance spectra with significantly improved statistical confidence. All timing-analysis results presented in this work are therefore derived from the combined signal of the seven stacked segments. Each individual observation shows features consistent with those seen in the stacked analysis, but with substantially lower signal-to-noise.

\subsection{Time-lag analysis}

The coherence between two light curves quantifies the degree to which their variability is correlated, or equivalently, how well the variations in one band can predict those in the other. A coherence value of 1 corresponds to perfectly coherent signals~\citep{Vaughan1997,Nowak1999,Vaughan2003,Uttley2014}. Fig.~\ref{freq_lag} (bottom panel) shows the coherence between the 0.3–0.7 keV (soft) and 0.9–10.0 keV (hard) bands. For each segment, we computed the Fourier transform and evaluated the coherence and its uncertainty following~\citet{Vaughan1997} and~\citet{Zoghbi2010}. The frequency bins were constructed so that the bin size is equal to 1.6 times the frequency value. As shown in Fig.~\ref{freq_lag}, the two bands remain highly coherent ($\geq 0.8$) up to $\sim3\times10^{-4}$ Hz. At higher frequencies, the coherence drops rapidly as poisson noise begins to dominate.

Fig.~\ref{freq_lag} (top panel) shows the time lag between the 0.3–0.7 keV and 0.9–10.0 keV bands as a function of Fourier frequency. The lags were computed following the method described in \citep{Nowak1999,Zoghbi2010,Uttley2014}, and we adopt the convention that a negative lag indicates that the soft-band variations follow those in the hard band. Between $8\times10^{-5}$ and $2\times10^{-4}$ Hz, the lags are negative, indicating that the hard-band variations lead. At frequencies below $8\times10^{-5}$ Hz, the lags are consistent with zero, although with relatively large uncertainties, and the finite duration of the observations prevents us from constraining the lag behavior at even lower frequencies. Consistent with the coherence results, the lag approaches zero at frequencies $\gtrsim4\times10^{-4}$ Hz, where the signal becomes dominated by poisson noise.

We also examine how the time lag varies as a function of energy within the frequency range $8\times10^{-5}$–$2\times10^{-4}$~Hz, where the hard-band variations lead those in the soft band. Based on the available photon statistics, we define a set of logarithmically spaced energy bins. For each bin, the lag is computed between the light curve in that bin and a broad 0.3–10 keV reference band. In constructing the reference band, the channel corresponding to the bin of interest is removed to ensure that the poisson noise remains uncorrelated~\citep{Uttley2014}. In this case, a negative lag indicates that variability in the given energy bin leads the reference band, whereas a positive lag indicates that it lags behind. The energy–lag spectrum obtained by stacking the seven observations is shown in Fig.~\ref{spectra}.

\subsection{Energy and covariance spectrum}
We fit the highest–signal-to-noise pn spectrum using XSPEC 12.14.0. The data were rebinned with \textsc{grppha} to ensure a minimum of 25 counts per bin, and the model was fit using $\chi^{2}$ statistics. The pn spectrum is modeled with \texttt{wabs*zwabs*(diskbb + zpowerlaw)}, consisting of a thermal disk–blackbody component and a nonthermal power-law component. Both components are subject to Galactic absorption as well as intrinsic absorption from the source. The best-fit model yields $\chi^{2}/{\rm dof} = 347/342 \simeq 1.0$. The photon index is $\Gamma = 2.67 \pm 0.09$, and the disk temperature is $kT = 0.14 \pm 0.01$ keV (Fig.~\ref{spectra}).

To investigate the origin of the correlated variability, we compute the covariance spectrum in the frequency range $8\times10^{-5}$–$2\times10^{-4}$~Hz. The covariance and its uncertainties are calculated following the method of~\citet{Wilkinson2009} and~\cite{Uttley2014}, using the same energy bins and reference band adopted for the energy–lag analysis. The resulting covariance spectrum can be described by a power law with a photon index of $\sim$2.0, which is flatter than the best-fitting photon index of the time-averaged spectrum. This shape indicates that the correlated variability is more strongly associated with the nonthermal emission component than with the thermal emission (Fig.~\ref{spectra}).

\section{Discussion and Conclusion}
As shown in Fig.~\ref{lc} (panel b), the X-ray blackbody temperature of \src\ rises from $\sim$70 eV to nearly $\sim$200 eV as the X-ray luminosity approaches its peak. Such high temperatures pose a challenge to the commonly adopted interpretation that the soft X-ray emission in TDEs originates from the inner regions of a standard accretion disk~\citep{Saxton2020}, since for a black hole of mass $\sim$$10^{7} M_\odot$ even maximal spin cannot produce effective disk temperatures near $\sim$200 eV~\citep{Yao2022}.
However, temperatures of this order are consistent with those typically inferred for the soft X-ray excess in AGN \citep{Gierlinski2004,Crummy2006,Jin2012,Done2012}.
\citet{Yao2022} proposed that this apparent discrepancy can be resolved if the observed color temperature is significantly boosted above the effective disk temperature due to comptonization, such that a large color-correction factor ($f_{\rm c}$) raises the apparent inner-disk temperature $T_{\rm in}$ well beyond the maximum temperature predicted for a thin disk.

Our timing analysis indicates that disk–corona reprocessing likely plays a significant role in producing the soft excess in \src. In AGNs, soft-band lags are widely interpreted as signatures of coronal hard X-ray photons irradiating the inner accretion disk and being reprocessed into lower-energy emission. The resulting reflection spectrum can generate the observed soft excess, either through blended low-ionization line emission or a bremsstrahlung-like continuum arising from free–free processes, and naturally introduces a light-travel delay between the hard and soft bands.
In a TDE, the returning stellar debris rapidly circularizes and forms a geometrically thin accretion disk, where the magnetorotational instability generates turbulence that dissipates gravitational energy~\citep{Xu2025}. A fraction of this dissipated energy is transported vertically by buoyant magnetic fields into an optically thin, hot coronal layer above the disk. Magnetic reconnection in this region heats the electrons to high temperatures, producing a comptonizing plasma~\citep{Xu2025}. 
The correspondence between our observed soft lag and this well-established reverberation behavior suggests that the soft excess in \src\ may originate from the same disk–corona interaction. In this context, the measured soft lag of $\sim$500 s implies a light-travel distance of roughly 10 gravitational radii for a black hole mass of $10^{7} M_{\odot}$.

\citet{DeMarco2013} conducted a systematic study of X-ray reverberation lags in AGNs and showed that both the characteristic lag frequency and lag amplitude scale with the SMBH mass. For a black hole mass of $10^{7} M_{\odot}$, their relations predict a reverberation frequency of $\sim$$3\times10^{-4}$ Hz and a lag amplitude of $\sim$100 s. Both values are somewhat offset from the observationally inferred ranges for \src, that is, $\nu_{\mathrm{lag}} \sim 8\times10^{-5} - 2\times10^{-4}$ Hz and $\tau \sim 500$ s. Similar deviations were also reported in AT2018fyk and were interpreted as arising from reprocessing in an outflow~\citep{Zhangwd2022}. However, unlike AT2018fyk—which shows no soft excess in the covariance spectrum—\src\ exhibits a clear soft excess component, indicating that the soft emission participates coherently in the correlated variability and is more consistent with a disk-reverberation origin~\citep{Uttley2011}.
 Such deviations have been reported in various AGNs and are often attributed to changes in the disk–corona geometry, frequently accompanied by flux-state transitions, including variations in the coronal height, size, or illumination pattern~\citep{Kara2013a,Alston2014,Uttley2014}.
A plausible explanation for this discrepancy in \src\ is an abrupt reduction in the magnetic power feeding the corona, such that when the magnetic energy injection weakens, the innermost coronal plasma is rapidly cooled by soft disk photons, causing the hard-X-ray–emitting region to appear more distant from the disk in terms of its illumination on short timescales. Such a geometric reconfiguration naturally produces a sudden drop in the observed X-ray luminosity, accompanied by a sharp increase in the spectral photon index and a marked decrease in the fraction of seed photons scattered into the corona ($f_{\rm sc}$), consistent with the parameter changes observed during the $\sim$3-day decline of \src\ after its peak (Fig.~\ref{lc}). Unfortunately, no \xmm\ observations were available during the peak-flux epoch of \src, preventing a direct test of this sudden change in the disk–corona geometry. Similar time-dependent geometric transitions have so far been clearly detected only in X-ray binaries~\citep{Kara2019,You2021}, but future improvements in observational capabilities may enable their identification in SMBH systems as well.

\begin{acknowledgements}
This research made use of the HEASARC online data archive services, supported by NASA/GSFC, and is based on observations obtained with \textit{XMM-Newton}. The author would like to thank Xinwen Shu, Chichuan Jin and Mingjun Liu for helpful discussions. This work is supported by the National Natural Science Foundation of China (grant no. 12433005, 12333004). 

\end{acknowledgements}

\bibliographystyle{bibtex/aa} 
\bibliography{ehb}

\end{document}